\documentclass[reprint,superscriptaddress,amssymb,amsmath,aps,10pt,floatfix,prb,longbibliography]{revtex4-2}
\usepackage{graphicx}
\usepackage{amsmath}
\usepackage{color,soul}

\usepackage{graphicx}

\usepackage{color}
\usepackage[colorlinks,bookmarks=false,citecolor=darkblue,linkcolor=red,urlcolor=blue]{hyperref}

\definecolor{darkred}{rgb}{0.7,0.0,0.0}
\definecolor{darkblue}{rgb}{0,0.02,0.45}

\begin{document}

\title{Antiferromagnetic resonance in  the cubic iridium hexahalides (NH$_4$)$_2$IrCl$_6$ and  K$_2$IrCl$_6$}

\author{L.~Bhaskaran}
\email{l.bhaskaran@hzdr.de}
\affiliation{Dresden High Magnetic Field Laboratory (HLD-EMFL) and
W\"urzburg-Dresden Cluster of Excellence ct.qmat, Helmholtz-Zentrum Dresden-Rossendorf, 01328 Dresden, Germany}

\author{A.~N.~Ponomaryov}
\thanks{Present Address: Institute of Radiation Physics, Helmholtz-Zentrum Dresden-Rossendorf, 01328 Dresden, Germany.}
\affiliation{Dresden High Magnetic Field Laboratory (HLD-EMFL) and    W\"urzburg-Dresden Cluster of Excellence ct.qmat, Helmholtz-Zentrum    Dresden-Rossendorf, 01328 Dresden, Germany}

\author{J.~Wosnitza}
\affiliation{Dresden High Magnetic Field Laboratory (HLD-EMFL) and W\"urzburg-Dresden Cluster of Excellence ct.qmat, Helmholtz-Zentrum  Dresden-Rossendorf, 01328 Dresden, Germany}
\affiliation{Institut f\"{u}r Festk\"{o}rper- und Materialphysik, TU Dresden, 01062 Dresden, Germany}

\author{N.~Khan}
\author{A.~A.~Tsirlin}
\affiliation{Experimental Physics VI, Center for Electronics Correlations and Magnetism, Institute of Physics, University of Augsburg, 86135 Augsburg, Germany}

\author{M.~E.~Zhitomirsky}
\affiliation{University Grenoble Alpes, Grenoble INP, CEA, IRIG, PHELIQS, 38000 Grenoble, France}

\author{S.~A.~Zvyagin}
\email{s.zvyagin@hzdr.de}
\affiliation{Dresden High Magnetic Field Laboratory (HLD-EMFL) and W\"urzburg-Dresden Cluster of Excellence ct.qmat, Helmholtz-Zentrum Dresden-Rossendorf, 01328 Dresden, Germany}

\date{october 16, 2021}%

\begin{abstract}
We report on high-field electron spin resonance studies of two iridium hexahalide compounds, (NH$_4$)$_2$IrCl$_6$ and K$_2$IrCl$_6$.  In the paramagnetic state,  our measurements reveal isotropic $g$-factors $g = 1.79$(1) for the Ir$^{4+}$ ions, in agreement with  their cubic symmetries. Most importantly,  in the magnetically ordered state, we observe two magnon  modes with zero-field gaps of 11.3 and 14.2~K for (NH$_4$)$_2$IrCl$_6$ and K$_2$IrCl$_6$, respectively. Based on that and using linear spin-wave theory, we estimate the nearest-neighbor exchange couplings and anisotropic Kitaev interactions, $J_1/k_B=10.3$~K, $K/k_B=0.7$~K for (NH$_4$)$_2$IrCl$_6$, and $J_1/k_B=13.8$~K, $K/k_B=0.9$~K for K$_2$IrCl$_6$, revealing  the nearest-neighbor Heisenberg coupling as the leading interaction term, with only a weak Kitaev anisotropy.
\end{abstract}

\maketitle
Transition-metal compounds with partly filled $4d$ and $5d$ orbitals have recently attracted enormous attention as promising candidates for realizing novel magnetic phenomena, emerging from the interplay between crystal-field effects and spin-orbit interactions~\cite{witczak2014correlated, rau2016spin, cao2018challenge}. In particular, honeycomb magnets with $j\rm_{eff}=\frac12$ moments and bond-anisotropic Kitaev interactions have come under extensive scrutiny due to the  possible realization of  quantum spin liquid states, characterized by unusual spin dynamics with fractionalized Majorana fermions and flux excitations~\cite{kitaev2006anyons, kitaev2006anyons, baskaran2007exact, jackeli2009mott, petrova2013unpaired, ponomaryov2020nature}.

While the majority of experimental and theoretical studies remains focused on honeycomb materials as the most promising platform for the Kitaev interaction,~\cite{takagi2019,winter2017models,tsirlin2021}, there is also a significant interest in detecting anisotropic exchange terms (including the Kitaev interaction) in Ir$^{4+}$ compounds with structures distinct from the honeycomb geometry~\cite{becker2015spin, cook2015spin, trebst2017kitaev, aczel2016, aczel2019revisiting, revelli2019spin}. However, most of the nonhoneycomb iridates studied over the last decade still evade detailed spectroscopic characterization with an unambiguous experimental determination of their interaction parameters. In particular, the absence of large single crystals as well as the notoriously strong neutron absorption by Ir atoms (which is of crucial importance for inelastic neutron scattering studies) and the insufficient energy resolution (in case of resonant inelastic x-ray scattering, RIXS) have often hindered the experimental access to relevant magnetic excitations.

\begin{figure}[t]
	\centering
	\includegraphics[width=0.47\textwidth]{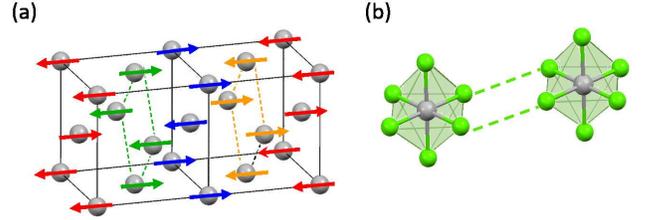}
	\caption{(a) Schematic view of the type-III antiferromagnetic spin structure of the IrCl$_6$ fcc lattice with four sublattices. Spins belonging to different sublattices are shown in red, green, blue, and orange colors. (b) Schematic representation of nearest-neighbor Cl$^{-}$-mediated superexchange pathways across the parallel edges of the octahedra formed by chlorine atoms. The Ir ions are shown in gray while the Cl ions are in green color.}
	\label{Molecule}
\end{figure}

Here, we focus on two compounds from the family of Ir$^{4+}$
antifluorites~\cite{griffiths1954complex, griffiths1959exchange, cooke1959exchange, harris1965magnetic}, (NH$_4$)$_2$IrCl$_6$ and K$_2$IrCl$_6$, that have been recently revisited by crystallographic and thermodynamic studies~\cite{khan2019cubic, reig2020structural, khan2021}. These compounds feature Ir$^{4+}$ ions arranged on a face-centered cubic~(fcc) lattice with four sublattices [Fig.~{\ref{Molecule}(a)}] and distinct long-range superexchange pathways mediated by the covalently bonded halogen ions [Fig.~\ref{Molecule}(b)]. The presence of possible frustration in fcc lattices is interesting in its own right as it shows an intricate competition of magnetic ground states, even in the absence of Kitaev or any other exchange anisotropy~\cite{lines1963antiferromagnetism, Schick20, balla2020}. In contrast to many other iridates, these compounds  are available as relatively large single crystals amenable to lab spectroscopic probes, featuring a cubic symmetry of the molecule that not only ensures the spin-orbit-coupled $j_{\rm eff}=\frac12$ state of the Ir$^{4+}$ ions, but also restricts the number of relevant interaction parameters. 

In the following, we present electron spin resonance (ESR) studies of (NH$_4$)$_2$IrCl$_6$ and K$_2$IrCl$_6$, revealing the presence of two gapped magnon modes in each of the compounds. Employing linear spin-wave-theory calculations, we estimate the spin-Hamiltonian parameters and, consequently, establish the main interaction terms.

Single crystals of (NH$_4)_2$IrCl$_6$ and K$_2$IrCl$_6$ were grown from solutions prepared by dissolving commercially available powders (Alfa Aesar) in deionized water and keeping the solutions at 60~$^{\circ}$C for several days. The resulting crystals had well-defined hexagonal (111) faces and sizes  of $\sim 2-3$ mm$^3$. Magnetization measurements revealed magnetic ordering transitions at 2.15~K for (NH$_4)_2$IrCl$_6$ and 3.1~K for K$_2$IrCl$_6$, in agreement with previous studies ~\cite{bailey1959specific, cooke1959exchange, reig2020structural,khan2019cubic}.

High-field electron spin resonance (ESR) measurements were performed employing a  transmission-type ESR spectrometer (similar to that described in Ref.~\cite{zvyagin2004high}), in magnetic fields up to 16 T. Measurements were done in the  frequency range of $60-400$~GHz, using a set of VDI microwave-chain radiation sources (product of Virginia Diodes, Inc., USA). An InSb hot-electron bolometer (QMC Instruments Ltd., UK) was used to record the spectra. We measured single crystals in the Faraday configuration with magnetic field $H$ $\parallel$ [111] and in the Voigt configuration with $H$ $\perp$ [111]. For the powder samples, we confined the specimens  by a binding agent such as high vacuum grease. The powders were obtained by crushing single crystals to ensure consistency between single-crystal and powder measurements. The stable free-radical molecule DPPH (2,2-diphenyl-1-picrylhydrazyl)  was used as a frequency-field marker.

\begin{figure}[t]
    \centering
    \includegraphics[scale=0.58]{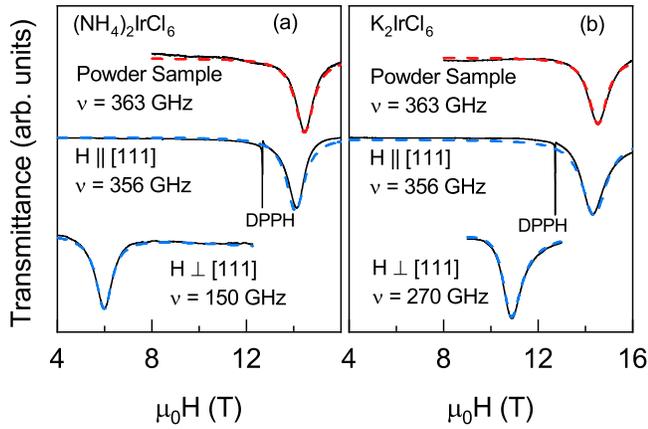}
    \caption{Normalized ESR spectra measured in single-crystalline ($H$ $\parallel$ $[$111$]$ and $H$ $\perp$ $[$111$]$) and powder samples of (a) (NH$_4$)$_2$IrCl$_6$ and  (b) K$_2$IrCl$_6$  at 50 K. The spectra are offset for clarity. Lorentzian fits to the powder sample are shown by the dashed red lines, while simulation of the single-crystal spectra are shown by the dashed blue lines.}
    \label{paraNH4-K2}
\end{figure}

ESR is traditionally recognized as a powerful tool to probe crystal-field effects in solids. A very sensitive test of the presence of non-cubic distortions in covalently bound [IrCl$_6$]$^{2-}$ complexes is provided by a $g$-factor anisotropy ~\cite{griffiths1954complex}. In  Fig. \ref{paraNH4-K2}, we show powder and single-crystal spectra  obtained for (NH$_4$)$_2$IrCl$_6$ (a) and K$_2$IrCl$_6$ (b) at a temperature of 50~K. Both powder spectra were fit using a Lorentzian function with   $g = 1.79(1)$, and linewidths of  0.93 and 1.23~T for (NH$_4$)$_2$IrCl$_6$ and K$_2$IrCl$_6$, respectively. These parameters perfectly agree with the simulation results for the single-crystals along both $H$ $\parallel$ [111] and $H$ $\perp$ [111], revealing  isotropic paramagnetic $g$-factors, and thus evidencing the cubic symmetry for both compounds. The reduction of the $g$-factor from the free-spin value ($g=2$) can be attributed to the covalent nature of the Ir and Cl bonds \cite{stevens1953magnetic,thornley1968magnetic}.
Large covalency of the Ir$^{4+}$ halides was indeed confirmed by recent RIXS studies~\cite{khan2021}, detecting also  minor deviations from the $j_{\rm eff}=\frac12$ state~\cite{reig2020structural}, despite the absence of any macroscopic symmetry lowering~\cite{khan2019cubic}.

\begin{figure}[t]
    \centering
    \includegraphics[scale=0.35]{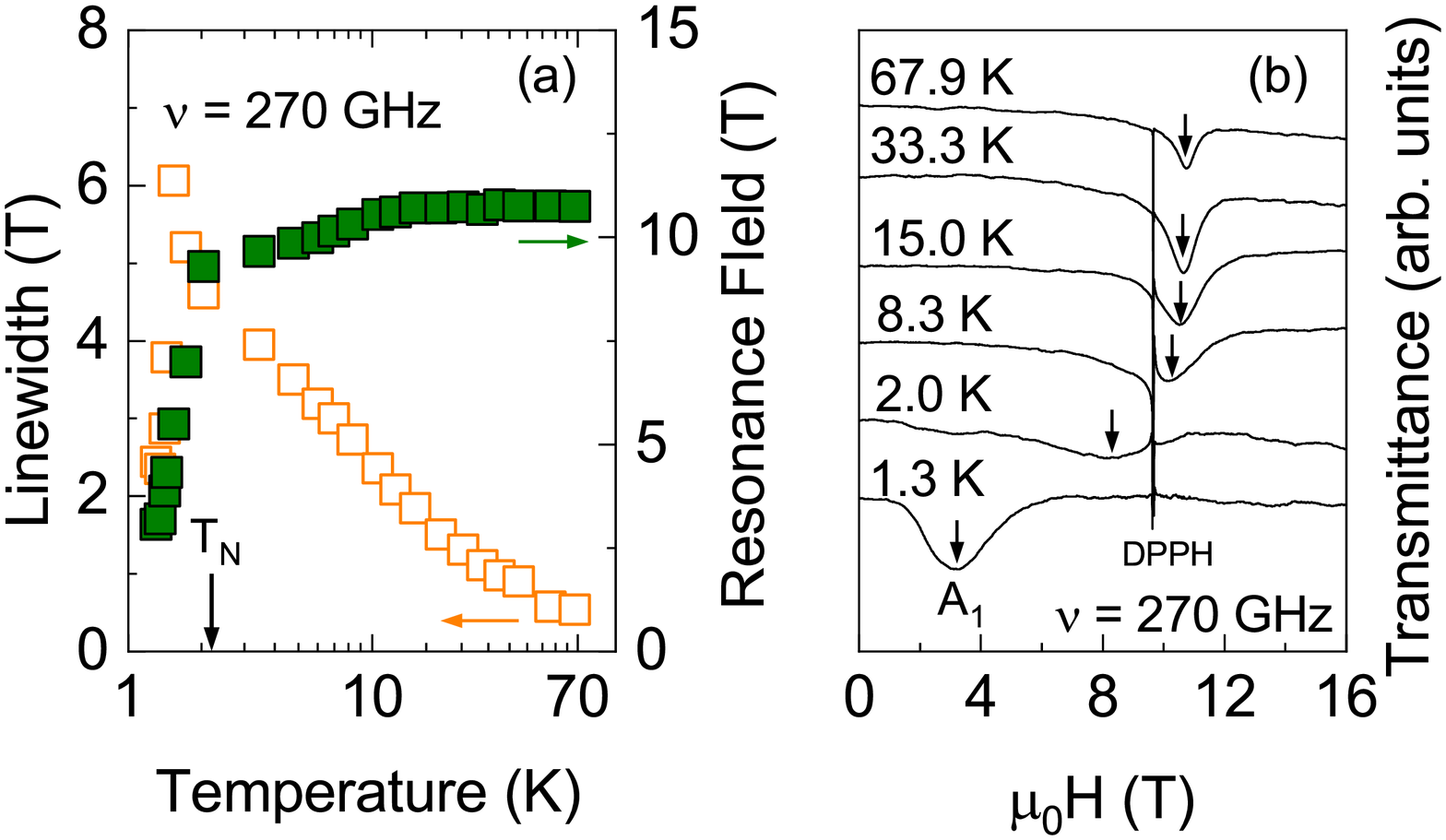}
    \caption{(a) Temperature dependence of the resonance field (closed squares) and linewidth (open squares) measured  in  (NH$_4$)$_2$IrCl$_6$ at 270 GHz. (b) Examples of normalized ESR spectra recorded at 270 GHz and different temperatures. The spectra are offset for clarity. Magnetic field is applied along the [111] direction.}
    \label{TDNH4}
\end{figure}

\begin{figure}[t]
    \centering
    \includegraphics[scale=0.35]{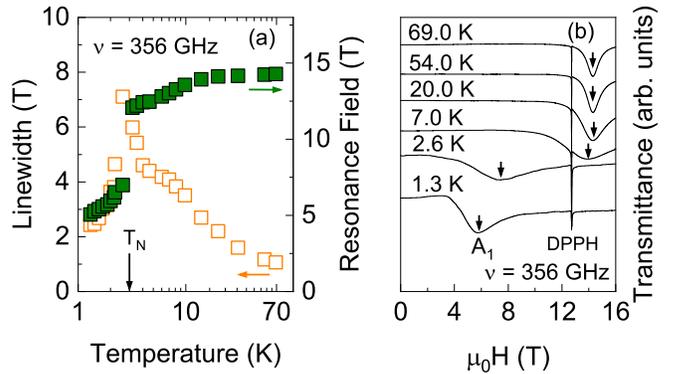}
    \caption{(a) Temperature dependence of the resonance field (solid squares) and linewidth (open squares) measured in K$_2$IrCl$_6$ at 356 GHz. (b) Examples of normalized ESR spectra recorded at 356 GHz and different temperatures. The spectra are offset for clarity. Magnetic field is applied along the [111] direction.}
    \label{TDK2}
\end{figure}

Upon cooling, the ESR spectra for $H$ $\parallel$ [111]  exhibit significant broadening,  evidencing  enhancement of magnetic correlations, with the maximum in the vicinity of  $T_N$ (Figs. \ref{TDNH4} and \ref{TDK2}). On further decreasing the temperature, the ESR linewidth rapidly decreases, revealing the onset of  long-range ordering below $T_N$. This transition into the ordered state   also results in a pronounced low-temperature shift of the ESR field positions.

\begin{figure}[t]
	\centering
	\includegraphics[scale=0.45]{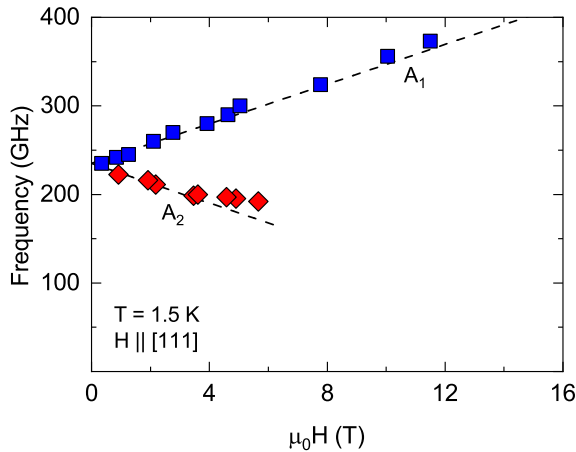}
	\caption{Frequency-field diagram of ESR excitations in (NH$_4$)$_2$IrCl$_6$ measured at 1.5 K  with magnetic field applied parallel to the $[$111$]$ direction. The dashed lines represent linear fits using Eq.~\eqref{FvsHfit}.}
	\label{FvsHNH4}
\end{figure}

\begin{figure}[t]
	\centering
	\includegraphics[scale=0.45]{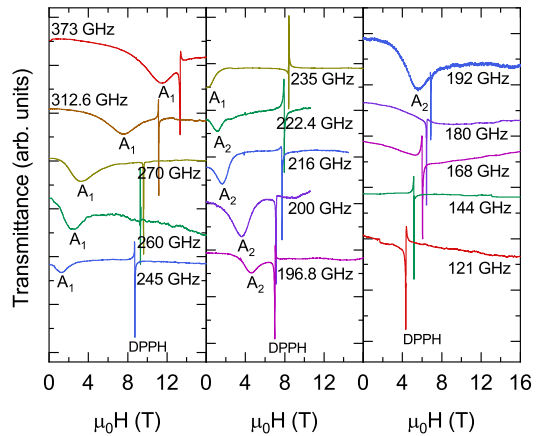}
	\caption{Examples of normalized ESR spectra for (NH$_4$)$_2$IrCl$_6$ at  different frequencies and $T=1.5$ K with $H\parallel$~[$111$].  The spectra are offset for clarity.}
	\label{SpectraNH4}
\end{figure}

\begin{figure}[t]
	\centering
	\includegraphics[scale=0.45]{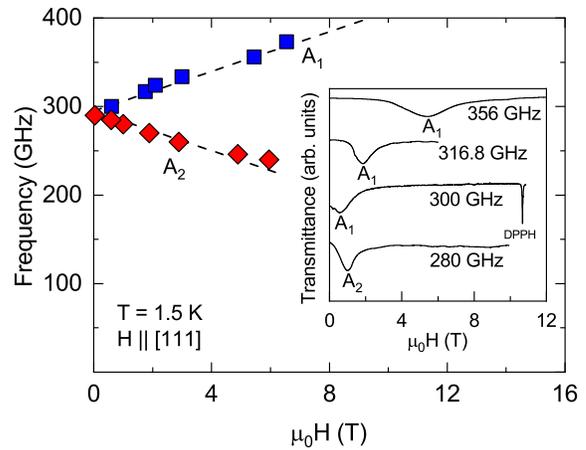}
	\caption{Frequency-field diagram of ESR excitations in K$_2$IrCl$_6$ measured at 1.5 K  with magnetic field applied parallel to the $[$111$]$ direction. The dashed lines represent linear fits using Eq.~\eqref{FvsHfit}. In the inset, examples of normalized ESR spectra for K$_2$IrCl$_6$ at different frequencies and $T=1.5$ K  with $H$ $\parallel$ $[$111$]$ are shown. The spectra are offset for clarity.}
	\label{FvsHK2}
\end{figure}

Two antiferromagnetic resonance (AFMR) modes, A$_1$ and A$_2$, were observed in  the magnetically ordered state. (Figs.~\ref{FvsHNH4} and \ref{FvsHK2}). Selected AFMR spectra measured at 1.5~K with magnetic field applied along the [111] direction are shown in Fig.~\ref{SpectraNH4} for (NH$_4)_2$IrCl$_6$  and in the inset of Fig. \ref{FvsHK2} for  K$_2$IrCl$_6$.

The frequency-field dependences of the modes A$_1$ and A$_2$  can be described using the equation
\begin{equation}
h\nu = \Delta \pm g_{eff} \mu_B B,
\label{FvsHfit}
\end{equation}
where $h$ is the Planck constant, $\nu$ represents the excitation frequency, $g_{eff}$ is the effective $g$-factor,   and $\mu_B$ the Bohr magneton. From this linear fit we obtain the zero-field magnon gaps $\Delta$ = 235 $\pm$ 10 and 295 $\pm$ 10~GHz (which  correspond to 11.3 and 14.2~K; $g_{eff} = 0.8$) for (NH$_4$)$_2$IrCl$_6$ and K$_2$IrCl$_6$, respectively. For both materials, mode A$_1$ shows a  linear increase in the resonance-field position, while some deviation from the linear dependence  was observed for mode A$_2$ in high fields. Noticeably, the mode A$_2$  disappears  above  $\sim 6$~T. This field corresponds to a field-induced transition reported for K$_2$IrCl$_6$~\cite{meschke2001field}, whereas in (NH$_4)_2$IrCl$_6$ the changes in applied field appear to be limited to a domain repopulation wherein the number of domains with magnetic moments oriented perpendicular to the applied field increases~\cite{khan2022}.



A general form of  spin-spin interactions in magnetic systems with an fcc crystal structure was discussed by several authors~\cite{Halg86,cook2015spin,kimchi2014kitaev,khan2019cubic}. Apart from the Heisenberg exchange it includes symmetry-allowed anisotropic terms typical for materials with strong spin-orbit coupling. According to an \textit{ab-initio} calculation~\cite{khan2019cubic} two strongest spin-spin interactions in (NH$_4$)$_2$IrCl$_6$ and K$_2$IrCl$_6$ are the Heisenberg exchange, $J_1>0$, and the Kitaev interaction, $K>0$, for the nearest-neighbor (nn) pairs of Ir ions. In addition, a weak next-nearest-neighbor (nnn) exchange $J_2>0$ is necessary to explain the experimentally observed type-III antiferromagnetic order~\cite{Hutchings1967}. Accordingly, we consider the following
Hamiltonian for effective $S=1/2$ moments of iridium ions:
\begin{equation}
\hat{\cal H} = \sum_{\langle ij\rangle}^{\rm nn} \Bigl( J_1 {\bf S}_i\cdot{\bf S}_j + K S_i^\gamma S_j^\gamma\Bigr)
+ J_2 \sum_{\langle ij\rangle}^{\rm nnn} {\bf S}_i\cdot{\bf S}_j \,, 
\label{HK}
\end{equation}
where ${\bf S}_i$ and  ${\bf S}_j$ are spin-1/2 operators at site $i$ and $j$,
respectively. In the anisotropic Kitaev term, $\gamma$ is one of the cubic axes ($x,y,z$) that is orthogonal to a given bond.

The Heisenberg fcc antiferromagnet with only the nearest-neighbor interactions  has highly degenerate classical ground states~\cite{Anderson50}. A small positive next-nearest-neighbor exchange $J_2$ lifts this degeneracy in favor of the type-III magnetic structure
described by the propagation vector ${\bf Q} = (2\pi/a,\pi/a,0)$~\cite{Yamamoto72} (note, that ${\bf Q}' = (0,\pi/a,2\pi/a) = - {\bf Q}+
{\bf G}$, where $\bf G$ is a reciprocal lattice vector).
In addition, the anisotropy  $K>0$ locks magnetic moments in the $(0,1,0)$ direction:
\begin{equation}
{\bf S}_i = \sqrt{2}\, S\hat{\bf y} \cos \bigl( {\bf Q}\cdot{\bf r}_i \pm \pi/4\bigr)\,.
\end{equation}
Together with  two other equivalent ${\bf Q}$, obtained by permutation of $Q_{x,y,z}$,  there are in total six domains for the type-III collinear structure.

Following a recent theoretical study of the excitation spectra in fcc antiferromagnets~\cite{Schick20}, we have performed linear spin-wave calculations for the Hamiltonian (\ref{HK}) with $J_2,K>0$. The ESR response  corresponds to long wave-length excitations with ${\bf k}\to 0$. Accordingly, the ESR spectrum is formed by two degenerate pairs of gapped magnons with the zero-field excitation energies:
\begin{equation}
\Delta_l = 2\sqrt{KJ_2}\,, \quad
\Delta_u = 2\sqrt{4KJ_1 + 2K^2 + KJ_2}\,.
\label{magnongap}
\end{equation}
Note, that the size of the low-energy gap $\Delta_l$ is determined by both $K$ and $J_2$, while the upper gap $\Delta_u$ depends mainly on the strength of the Kitaev interaction.  Moreover, in the absence of $K$ both gaps vanish, while at $J_2\to 0$ they have substantially different magnitudes. 

The next-nearest-neighbor isotropic exchange coupling was recently estimated from the \textit{ab initio} calculation
as $J_2/k_B\simeq 0.2$~K for both (NH$_4$)$_2$IrCl$_6$ and K$_2$IrCl$_6$~\cite{khan2022,khan2019cubic}. 
In addition, we use the expression for the  Curie-Weiss temperature \cite{ashcroft1976solid}
\begin{equation}
 \theta_{\rm CW}=3J_1+K+\frac32\,J_2 \ .
 \label{CurieWeiss}
\end{equation}

Experimental values for  $\theta_{\rm CW}/k_{B}$ are 32 K for (NH$_4$)$_2$IrCl$_6$~\cite{khan2022} and 42.6~K for K$_2$IrCl$_6$~\cite{khan2019cubic}. They are obtained by fits to the magnetic-susceptibility data in the 100-400 K and 150-400 K range, respectively.  The corresponding effective moment of 1.73 $\mu_B$ confirms that mostly the ground-state $j_{\rm eff}=\frac12$ doublet is occupied in this temperature range. Using Eq.~\eqref{magnongap} and Eq.~\eqref{CurieWeiss} we estimate $J_1/k_B=10.3$~K and $K/k_B=0.7$~K for  (NH$_4$)$_2$IrCl$_6$, and $J_1/k_B=13.8$~K and $K/k_B=0.9$~K for K$_2$IrCl$_6$ ($K/J_1\approx 6-7\%$).  The exchange coupling $J_1$ is stronger for K$_2$IrCl$_6$ compared to (NH$_4$)$_2$IrCl$_6$, due to the lattice parameter difference ($a=9.77$\,\r A vs.\ 9.87\,\r A~\cite{khan2019cubic, khan2022}, respectively), with the respective negative pressure reducing magnetic interactions in (NH$_4)_2$IrCl$_6$. It is worth noting that our experimental value of $K$ is much smaller than the 5 K reported in Ref.~\cite{khan2019cubic}. This discrepancy may indicate the importance of Hund's coupling on the Cl ligand, an effect that has been neglected in the superexchange theory used in Ref.~\cite{khan2019cubic} and generally leads to a ferromagnetic contribution, which counteracts antiferromagnetic superexchange. Similar effects were previously reported in Cu$^{2+}$ halides~\cite{lebernegg2013magnetism}.

Using the parameters, as obtained above, the size of low-energy gaps was also estimated, yielding $\Delta_l/k_B\approx 1$~K.  Our thorough search  did not reveal the low-energy AFMR modes (note also that our ESR probe has a  cut-off frequency limit of 60-70 GHz that is   higher than the expected size of $\Delta_l$). The absence of the low-energy ESR modes, which directly depends on $J_2$, may indicate, on the other hand,  that the next-nearest-neighbor exchange in these compounds is very small. In such a case the quantum order-by-disorder effect can play a significant role in the Kitaev fcc antiferromagnets \cite{li2017kitaev}. 

Summarizing, we presented systematic ESR studies of the two iridium hexahalide compounds (NH$_4$)$_2$IrCl$_6$ and K$_2$IrCl$_6$. In the paramagnetic state, our measurements reveal isotropic $g$-factors $g = 1.79$(1) for Ir$^{4+}$ ions, confirming  their cubic symmetries. By measuring the ESR spectra in the ordered state, we were able to determine the nearest-neighbor Heisenberg exchange $J_1$ and the Kitaev interaction $K$ for both materials, revealing  $K\ll J_1$. The intriguing possibility of the realization of the  quantum order-by-disorder effect as well as details of the high-field behavior of magnetic excitations  in these frustrated compounds deserves further experimental and theoretical studies. 

This work was supported by the Deutsche Forschungsgemeinschaft (DFG), through ZV 6/2-2, 
the W\"urzburg-Dresden Cluster of  Excellence on Complexity and Topology in Quantum Matter–$ct.qmat$ (EXC 2147, Project No. 390858490),  and SFB 1143, as well as by the HLD at HZDR, member of the European Magnetic Field Laboratory (EMFL). The work in Augsburg was supported by the DFG via the Project No. 107745057 (TRR80). M.E.Z.\ acknowledges financial support from ANR, France, Grant No. ANR-18-CE05-0023.

\bibliography{IrCl6v32}
\end{document}